\documentclass[
	a4paper, 
	10pt,    
	twoside, 
]{LTJournalArticle}

\usepackage{amssymb}
\usepackage{amsmath}

\usepackage{float}

\runninghead{Passage of a GRB Through a Molecular Cloud} 

\footertext{\textit{Astronomy Letters} (2024) 50:510-522} 

\title{Passage of a Gamma-Ray Burst Through a Molecular Cloud:\\ The Absorption of its Afterglow in the X-ray Wavelength Range} 

\author{%
	A.V. Nesterenok\textsuperscript{1}\thanks{Corresponding author: \href{mailto:alex-n10@yandex.ru}{alex-n10@yandex.ru}\\ \textbf{Received:} August 14, 2024, \textbf{Published:} August, 2024}
}

\date{\footnotesize\textsuperscript{\textbf{1}}Ioffe Institute, Saint Petersburg, Russia}

\renewcommand{\maketitlehookd}{%
	\begin{abstract}
	\normalsize \rm
	\noindent We study the absorption of a gamma-ray burst afterglow in a dense molecular cloud in the X-ray wavelength range. We report the results of numerical simulations of the propagation of the gamma-ray burst radiation in the cloud for various gas densities, metallicities, and distances from the gamma-ray burst progenitor star and the cloud. We consider a sample of 45 gamma-ray bursts with known redshifts in which the isotropic-equivalent gamma-ray energy is approximately equal to the value adopted in our numerical simulations.
	For these gamma-ray bursts, we have analyzed the Swift/XRT energy spectra of their afterglows at late times, $t \geq 4 \times 10^3$~s. It is shown that the hydrogen column densities estimated using the absorption model in which the ionization of metal ions is not taken into account and the solar metallicity is used are a factor of 1--3 smaller than the actual values -- if the molecular cloud is located close to the gamma-ray burst progenitor star. If the gas--dust cloud is located at a distance of $R \geq 10$~pc from the source of the gamma-ray burst or the gas metallicity is $[M/H] \leq -1$, then the effect of the cloud ionization structure on the absorption of the afterglow is minor.    
	
	\end{abstract}
}

\begin{document}
\maketitle 

\section{Introduction}
The gamma-ray bursts result from the explosion and the release of a huge amount of gamma-ray radiation energy, up to $\sim 10^{52}$~erg or even higher, into the surrounding space \autocite{Goldstein2016}. Owing to the enormous energy release, gamma-ray bursts play a pivotal role in the study of the present-day astrophysical problems (see, e.g., \autocite{Bozzo2024}).
The classification of gamma-ray bursts is based on the properties of the emission in the first, prompt phase -- the gamma-ray emission with a set of overlapping peaks. The duration of the gamma-ray burst prompt emission is usually defined as the time interval during which the fluence increases from 5 to 95 per cent. This parameter is denoted by $T_{90}$ and varies for gamma-ray bursts from fractions of a second to several thousand seconds.
Gamma-ray bursts are divided into two main classes -- short ($T_{90} \lesssim 2$~s) and long ($T_{90} \gtrsim 2$~s). These two classes of gamma-ray bursts occupy two well-separated regions on the diagram spectral hardness ratio -- prompt phase duration \autocite{Pozanenko2021}. Until recently, it was generally believed that long gamma-ray bursts are produced through the collapse of massive stars, while short gamma-ray bursts result from the merger of two compact stars.
The simultaneous detection of the short gamma-ray burst GRB~170817A and the gravitational wave signal GW~170817 provided a strong evidence that short gamma-ray bursts are associated with the compact star merger events \autocite{Goldstein2017}. And the recent discovery of kilonova emission in the afterglows of some long gamma-ray bursts forced one to revise the generally accepted view of the origin of long gamma-ray bursts \autocite{Troja2022,Levan2024}.

An important achievement in the study of gamma-ray bursts was the discovery of an afterglow in the optical and X-ray wavelength ranges -- this was done for GRB~970228 \autocite{vanParadijs1997,Costa1997}. And for GRB~970508, the redshift was estimated for the first time from absorption lines in the optical spectrum of the afterglow: $0.835 < z \lesssim 2.3$ \autocite{Metzger1997}.  
This discovery confirmed the hypothesis about the cosmological origin of gamma-ray bursts. The redshift of a gamma-ray burst can also be determined if it is possible to localize the gamma-ray burst in the sky and to identify its host galaxy \autocite{Kulkarni1998}. The gamma-ray burst afterglow undergoes absorption in the interstellar medium of the host galaxy. The afterglow energy spectrum with many absorption lines of metal ions, hydrogen atoms and molecules contains information about the interstellar medium of the galaxy \autocite{Schady2017}. The redshift of the most distant gamma-ray bursts detected to date is $z \approx 8-9$ (see, e.g., \autocite{Cucchiara2011,Tanvir2018}).

An analysis of the spectral flux density of a gamma-ray burst afterglow in the infrared and optical wavelength ranges allows one to determine the visual extinction $A_{\rm V}$ in the host galaxy. About 30 per cent of gamma-ray burst afterglows have no evidence of absorption in the continuum in the optical wavelength range, $A_{\rm V} \lesssim 0.1$. At the same time, about 25 per cent of gamma-ray bursts are "dark", i.e., the emission in the optical wavelength range experiences strong absorption, $A_{\rm V} > 1$ \autocite{Covino2013}. 
It is possible to estimate the hydrogen column density $N_{\rm HX}$ through which the gamma-ray burst emission passed from the analysis of the afterglow energy spectrum in the X-ray wavelength range. It has been found that the hydrogen column density $N_{\rm HX}$ can exceed by an order of magnitude or more the hydrogen column density $N_{\rm H}$, evaluated from data in the optical wavelength range -- from atomic hydrogen lines, metal ion lines, or $A_{\rm V}$ \autocite{Watson2007,Campana2010,Schady2011}.
The origin of the absorption of the gamma-ray burst afterglow in the X-ray wavelength range -- the ionized or neutral gas, the location in the host galaxy or in the intergalactic medium -- remains unclear \autocite{Schady2017,Dalton2020}.

In the analysis of the afterglow energy spectrum in the X-ray wavelength range, it is usually assumed that the absorbing gas is neutral, has solar metallicity, and experiences no ionization effects from the gamma-ray burst radiation (i.e., it is located at a large distance from the source of the gamma-ray burst in the host galaxy).
Studies of host galaxies of gamma-ray bursts at redshifts $z \lesssim 1$ suggest that the gas metallicity in such galaxies is typically lower than the solar metallicity (see, e.g., \autocite{Vergani2015}). The theoretical models of the evolution of massive stars predict that in order to maintain a high rotation rate (it is necessary for the formation of a massive accretion disk and the launch of an ultra-relativistic jet during the collapse of star core), the metallicity of the star must be lower than the solar one, $Z < 0.3Z_{\odot}$ \autocite{Woosley2006}.
In addition, the intense radiation of the gamma-ray burst can ionize the interstellar gas at a distance of ten parsecs or more (see, e.g., \autocite{Watson2013,Krongold2013,Nesterenok2024}). Thus, the assumptions that are made in the evaluation of the parameter $N_{\rm HX}$ -- solar metallicity and an unperturbed interstellar medium -- may not hold.

The problem of the interaction of the gamma-ray burst radiation with the interstellar gas has been considered by many authors (see, e.g., \autocite{Perna2000,Draine2002,Perna2002,Barkov2005,Barkov2005b}). In our previous paper we presented a numerical model of the passage of the gamma-ray burst radiation through a dense gas--dust cloud \autocite{Nesterenok2024}. It was shown that the gas is completely ionized by the gamma-ray burst radiation at a distance of several parsecs.
At the same time, the boundary between the fully ionized gas and the neutral gas, where the ionization fraction is much less than unity, is sharp. It was shown by Nesterenok \autocite{Nesterenok2024} that in a layer of neutral gas located close to the ionization front, the metal ions have a high charge due to the ionization by X-rays. The farther from the ionization front, the higher the abundance of ions with a low charge.
Accordingly, one might expect that the contribution of metal ions to the absorption of radiation in the X-ray wavelength range will depend on the hydrogen column density in the gas--dust cloud. The aim of this paper is to investigate the effect of the cloud ionization structure on the absorption of the gamma-ray burst afterglow in the X-ray wavelength range. We use the results of our numerical simulations to fit energy spectra of gamma-ray burst afterglows obtained with the XRT telescope onboard the Swift observatory \autocite{Burrows2005}.

\section{The numerical model}
A detailed description of the numerical model for the passage of the gamma-ray burst radiation through a gas--dust cloud is presented in \autocite{Nesterenok2024}. The model takes into account the ionization of H and He atoms, the ionization of metal ions with the emission of Auger electrons, the ionization and photodissociation of H$_2$ molecules, the thermal sublimation of dust grains, and the absorption of ultraviolet (UV) radiation via Lyman and Werner band transitions of H$_2$ molecules. Below we briefly outline the main details of the numerical simulations and point out the changes made to the numerical model.

In the numerical model, the cloud is located at some distance $R_{\rm min}$ from the progenitor star of the gamma-ray burst. For most of our numerical calculations that will be discussed in this paper, the distance $R_{\rm min}$ was chosen to be 1~pc. In this paper we also discuss the results of our numerical simulations in which the cloud is located at a distance $R_{\rm min} = 10$ and 100~pc from the gamma-ray burst progenitor star.
Another parameter of the model is the total number density of hydrogen nuclei in the cloud $n_{\rm H,tot}$ -- the number density of hydrogen atoms in the molecular, atomic, or ionized state. The calculations were carried out for three values of the parameter $n_{\rm H,tot}$ -- $10^2$, $10^3$, and $10^4$~cm$^{-3}$. The gas molecular fraction is

\begin{equation}
f_{\rm H_2} = \frac{2n_{\rm H_2}}{n_{\rm H} + 2n_{\rm H_2}}.
\end{equation}

\noindent
At the initial time the gas is neutral, the gas molecular fraction $f_{\rm H_2}$ was taken to be 0.1 or 1. The chemical elements H, He, C, N, O, Ne, Mg, Si, S, and Fe are taken into account in the numerical model. The abundances of metal ions are equal to the abundances in the Solar System with a correction for the metallicity $[M/H] \equiv {\rm log_{10}} (Z/Z_{\odot})$.
The abundances of chemical elements in the case $[M/H] = 0$ are presented in Table~1 (Lodders et al. \autocite{Lodders2009}; table~4 in their paper). In our model we take into account the gamma-ray burst optical flash, prompt, and afterglow emission. The total isotropic equivalent energy of the prompt emission is $E_{\gamma, {\rm iso}} = 5 \times 10^{52}$~erg, while the peak energy of the spectral energy distribution $\nu F_{\nu}$ in the rest frame is $E_{\rm peak} = 350$~keV.
The efficiency of the gamma-ray burst prompt phase $\eta$ (the ratio of the energy radiated in the prompt phase $E_{\gamma, {\rm iso}}$ to the total explosion energy) is 0.17 in our model. The remaining parameters of the gamma-ray burst emission are given in \autocite{Nesterenok2024}.

\begin{table}  
	\caption{\normalsize Abundances of chemical elements relative to the hydrogen nuclei (for solar metallicity, $[M/H] = 0$)}
	\centering
	\begin{tabular}{l l}
	\toprule
	He & 0.0841 \\               
	C  & $2.46 \times 10^{-4}$ \\
	N  & $7.24 \times 10^{-5}$ \\
	O  & $5.37 \times 10^{-4}$ \\
	Ne & $1.12 \times 10^{-4}$ \\
	Mg & $3.47 \times 10^{-5}$ \\
	Si & $3.31 \times 10^{-5}$ \\
	S  & $1.38 \times 10^{-5}$ \\
	Fe & $2.82 \times 10^{-5}$ \\
	\bottomrule
	\end{tabular}
	\label{tab:distcounts}
\end{table}

The density of the dust material was chosen to be $\rho_{\rm d} = 3.5$~g/cm$^{3}$, and its chemical composition was taken to be MgFeSiO$_4$. The dust-to-gas mass ratio is determined by the metallicity and the dust depletion of metals. The dust depletion of Fe ions was chosen to be 0.99. The metal ions that are not locked in dust grains are in the gas phase. 
For metallicity $[M/H] = 0$, the dust-to-gas mass ratio is 0.0036 at the initial time. In contrast to the work \autocite{Nesterenok2024}, we took into account the cross sections for the absorption and scattering of photons by dust grains for photon energies $h \nu < 12$~keV, calculated in \autocite{Draine2003}. The data on the dust dielectric function are available on the website of Prof. Draine\footnote{\url{https://www.astro.princeton.edu/~draine/dust/dust.diel.html}, the file callindex.out\_silD03}. 
The scattering of photons by dust grains is not taken into account for photon energies $h \nu > 1$~keV \autocite{Draine2002}. In our model we consider the thermal sublimation of dust grains as a result of the heating by the optical and UV radiation of the gamma-ray burst; other destruction mechanisms of dust grains are disregarded \autocite{Draine2002,Lu2021}.

The cloud is divided into spherical shells. The inner radius of a shell $j$ is $R_j$, and the thickness $\Delta R$ is the same for all shells. The value of the parameter $\Delta R$ was chosen in the range $2.5 \times 10^{16} - 10^{17}$~cm depending on the gas density. The optical depth on the dust for the shell $\Delta R$ is much less than unity.
Sequentially for each cloud shell, the system of differential equations is solved for the energy level populations of H$_2$ molecule, the number densities of ions and chemical compounds, and the dust grain radius and temperature. The system of differential equations is solved using the numerical code SUNDIALS CVODE v5.7.0 \autocite{Hindmarsh2005,Gardner2022}. The result of the solution of the system of equations is the dependence of the variables on the retarded time $t_{r}$ for each cloud shell ($t_{r} = 0$ corresponds to the time moment when the gamma-ray burst radiation front reaches the cloud shell).
The parameters of our numerical models are provided in Tables~2 and 3. The models 1, 2 and 3 differ by the total number density of hydrogen nuclei in the cloud. The additional models 4 and 5 differ from other models in that the gas metallicity in them was taken to be $[M/H] = -0.5$ and $-1$, respectively. In models 6 and 7 the distance from the gamma-ray burst source to the cloud is $R_{\rm min} = 10$ and 100~pc, respectively. It is assumed that the gamma-ray burst emission does not experience any absorption on the way from the burst source to $R_{\rm min}$.

\begin{table*}
	\caption{\normalsize Basic numerical models}
	\centering 
	\begin{tabular}{p{10cm}|p{1cm}|p{1cm}|p{1cm}}
	\toprule
	Model & 1 & 2 & 3 \\
	\midrule
	Distance from the gamma-ray burst progenitor star, $R_{\rm min}$ $\left[\right.$pc$\left.\right]$ & 1 & 1 & 1 \\
	Gas density, $n_{\rm H,tot}$ $\left[\right.$cm$^{-3}\left.\right]$ & $\mathbf{10^2}$ & $\mathbf{10^3}$ & $\mathbf{10^4}$ \\
	Molecular fraction, $f_{\rm H_2}$ & {\bf 0.1} & 1 & 1 \\
	Metallicity, $[M/H]$ & 0 & 0 & 0 \\
	\bottomrule
	\end{tabular}
\end{table*}

\begin{table*}
	\caption{\normalsize Additional numerical models}
	\centering 
	\begin{tabular}{p{10cm}|p{1cm}|p{1cm}|p{1cm}|p{1cm}}
	\toprule
	Model & 4 & 5 & 6 & 7 \\
	\midrule
	Distance from the gamma-ray burst progenitor star, $R_{\rm min}$ $\left[\right.$pc$\left.\right]$  & 1 & 1 & {\bf 10} & {\bf 100} \\
	Gas density, $n_{\rm H,tot}$ $\left[\right.$cm$^{-3}\left.\right]$ & $10^3$ & $10^3$ & $10^3$ & $10^3$ \\
	Molecular fraction, $f_{\rm H_2}$ & 1 & 1 & 1 & 1 \\
	Metallicity, $[M/H]$ & $\mathbf{-0.5}$ & $\mathbf{-1}$ & 0 & 0 \\
	\bottomrule
	\end{tabular}
\end{table*}

The optical depth that radiation passes to the inner boundary of a cloud shell $j$ is the sum of the contributions to the optical depth from all of the cloud shells preceding the shell $j$:

\begin{equation}
\tau_{j-1}(E,t_r) = \sum_{i=1}^{j-1} \Delta \tau_i(E,t_r),
\end{equation}

\noindent
where $E$ is the photon energy, and $t_r$ is the retarded time, the same for all shells. The result of numerical simulations is a two-dimensional table of $\tau(E, N_{\rm H,tot})$ -- the optical depth as a function of the photon energy $E$ and the column density of hydrogen nuclei in the cloud $N_{\rm H,tot}$. The value of the parameter $N_{\rm H,tot}$ at a distance $R_j$ is

\begin{equation}
N_{\rm H,tot} = n_{\rm H,tot} (R_j - R_{\rm min}).
\end{equation}

\noindent
The step of the hydrogen column density grid is equal to the hydrogen column density in one cloud shell:

\begin{equation}
\Delta N_{\rm H,tot} = n_{\rm H,tot} \Delta R.
\end{equation}

\noindent
In our numerical simulations, the table of $\tau(E, N_{\rm H,tot})$ was saved for a retarded time $t_r = 10^5$~s. The spectral flux density of the radiation from the source is

\begin{equation}
F(E, N_{\rm H,tot}) = F_{\rm 0}(E) \exp\left(-\tau(E, N_{\rm H,tot})\right),
\end{equation}

\noindent
where $F(E, N_{\rm H,tot})$ is the power of the radiation per unit solid angle for photon energies $E$ after its passage through the gas--dust cloud.

The radiation of the gamma-ray burst is responsible for the complete ionization of the gas near the burst source. The boundary between the region where the gas is fully ionized and the region where the gas is predominantly neutral, has a size of $\approx 0.05$~pc for a gas density $n_{\rm H,tot} = 10^3$~cm$^{-3}$ \autocite{Nesterenok2024}.
Since the shell of fully ionized gas has almost no effects on the afterglow energy spectrum, we shift the reference point of the hydrogen column density to the ionized--neutral gas boundary:

\begin{equation}
N_{\rm HX} = N_{\rm H,tot} - N_0,
\label{eq_nhx_def}
\end{equation}

\noindent
where $N_0$ is the hydrogen column density of the fully ionized cloud shell. It is the values of the parameter $N_{\rm HX}$ that are determined from the analysis of energy spectra of gamma-ray burst afterglows. The values of $\tau(E, N_{\rm HX})$ calculated in our numerical simulations are written in the form of a two-dimensional table in the FITS format\footnote{A code written in the python programming language was used to write the data in the FITS format. The code is available at \url{https://github.com/mbursa/xspec-table-models} and is distributed under the MIT license.}.

\section{Observational data processing}
Tsvetkova et al. \autocite{Tsvetkova2017,Tsvetkova2021} published the catalogues of gamma-ray bursts with measured redshifts, that were observed with the Konus gamma-ray spectrometer onboard the Wind satellite (NASA). The first catalogue consists of 150 gamma-ray bursts (138 of them are long gamma-ray bursts) detected in the trigger mode in the Konus--Wind experiment from 1997 to June 2016 \autocite{Tsvetkova2017}. 
The second catalogue comprises 167 gamma-ray bursts (160 of them are long gamma-ray bursts) detected simultaneously with the BAT telescope onboard the Swift space observatory and the Konus gamma-ray spectrometer in the waiting mode from 2005 to 2018 \autocite{Tsvetkova2021}. Estimates of the total isotropic equivalent energy of the prompt emission of gamma-ray bursts are provided in the catalogues by \autocite{Tsvetkova2017,Tsvetkova2021}. From both catalogues we choose 45 long gamma-ray bursts for which the total isotropic equivalent energy $E_{\gamma, {\rm iso}}$ lies in the range from $3.33 \times 10^{52}$~erg to $7.5\times 10^{52}$~erg -- a factor of 1.5 lower and a factor of 1.5 higher than the value of $E_{\gamma, {\rm iso}}$ adopted in our numerical simulations.

In our paper we analyze energy spectra of the afterglows of selected gamma-ray bursts, that were obtained with the XRT onboard the Swift observatory \autocite{Burrows2005}. The energy range of the XRT/Swift is $0.2-10$~keV. The energy spectra of gamma-ray bursts were generated through an automatic processing of the observational data by the Swift team and are available on the Swift website\footnote{\url{https://www.swift.ac.uk/xrt_spectra/}} \autocite{Evans2009}.
For our analysis we took energy spectra obtained at late times $t \geq 4 \times 10^3$~s in the photon counting (PC) mode. In this case, the mean photon arrival time is typically $t > 10^4$~s. The spectral-fitting program XSpec of version 12.13.1 in the HEASoft\footnote{\url{https://heasarc.gsfc.nasa.gov/docs/software/heasoft/}} software packages was used to analyze the energy spectra of gamma-ray burst afterglows \autocite{Arnaud1996,Dorman2001}.  
The energy spectra were fitted with a power function by taking into account the absorption in the interstellar medium in the host galaxy at redshift $z$ and in our Galaxy. The spectra were fitted using the following composite models:

\begin{equation}
\begin{split}
& 1. \quad \textrm{tbabs * ztbabs * powerlaw}, \\
& 2. \quad \textrm{tbabs * etable\{model\_name.fits\} * powerlaw}, 
\end{split}
\label{eq_fit_models}
\end{equation}

\noindent
where {\bf tbabs} and {\bf ztbabs} are the T{\"u}bingen-Boulder models for the absorption of X-rays in the interstellar medium from \autocite{Wilms2000} at zero redshift and a given redshift $z$, respectively. The {\bf etable\{model\_name.fits\}} model uses a file with the extension FITS which contains the results of our numerical simulations in the form of a table.
The first component in each of the composite models in (\ref{eq_fit_models}) is responsible for the absorption of X-rays by the interstellar gas in our Galaxy. The only parameter of this model is the hydrogen column density $N_{\rm H,Gal}$. 
The values of $N_{\rm H,Gal}$ for each burst were determined using the online tool at the Swift website\footnote{\url{https://www.swift.ac.uk/analysis/nhtot/index.php}} \autocite{Willingale2013}. The second component in each of the composite models in (\ref{eq_fit_models}) is responsible for the absorption of X-rays in the host galaxy at a given redshift $z$. The last component in the models (\ref{eq_fit_models}) defines a power law spectrum.
We used the photoionization cross sections of metal ions published by Verner and Yakovlev \autocite{Verner1995} and Verner et al. \autocite{Verner1996}. The abundances of chemical elements were taken to be equal to those in the solar photosphere published by Lodders et al. \autocite{Lodders2009} -- this choice is made using the {\bf abund lpgp} command in XSpec.
The energy spectra were also fitted using the absorption models (at the redshift of the host galaxy) in which the gas metallicity was taken to be $[M/H] = -0.5$ and $-1$. The hydrogen column density $N_{\rm HX}$ in the host galaxy at a redshift $z$, the exponent, and the normalization factor of the energy spectrum are free parameters.  
In the spectrum fitting routine, it was assumed that the signal from the source and the background obey the Poisson distribution. The choice of this statistic in the XSpec program is specified by the command {\bf statistic cstat}. The errors of the parameters were evaluated using the {\bf error} command. The confidence regions of the parameters were derived using the $\chi^2$ statistic. The parameter {\bf delta fit statistic} was chosen to be 2.706, which corresponds to a confidence level of 90 per cent.

\section{The T{\"u}bingen-Boulder X-ray absorption model (tbabs)}
The absorption model from Wilms et al. \autocite{Wilms2000} describes the absorption of X-rays in the interstellar medium. It is assumed in their model that the absorbing gas is neutral, the gas molecular fraction $f_{\rm H_2}$ is 20 per cent. Note that the ratio of the photoionization cross sections of H$_2$ and H is 2.95 at photon energies $E = 1$~keV. 
The absorption model of Wilms et al. \autocite{Wilms2000} takes into account the chemical elements Na, Al, P, Cl, Ar, Ca, Ti, Cr, Mn, Co, and Ni (in addition to those that are considered in our model). The contribution of the photoionization of these metals to the optical depth due to the photoionization of more abundant metal ions (C, N, O, Ne, Mg, Si, S, Fe) is about 3 per cent at photon energies of 1~keV.
The model of Wilms et al. \autocite{Wilms2000} does not take into account the Compton ionization of H, H$_2$ and He that makes a major contribution to the optical depth at photon energies $\geq 10$~keV. However, the optical depth is small at such photon energies: $\tau < 0.1$ for hydrogen column densities $N_{\rm HX} < 10^{23}$~cm$^{-2}$. 
Another difference between the absorption model of Wilms et al. \autocite{Wilms2000} and our model is that in our model the chemical composition of the dust is MgSiFeO$_4$. In the model of Wilms et al. \autocite{Wilms2000}, the dust grains are composed of many more elements, with each element having its own dust depletion factor. The dust-to-gas mass ratio is 0.005 in the model of Wilms et al. \autocite{Wilms2000}.

\begin{figure*}
\centering
\includegraphics[width = 1.\textwidth]{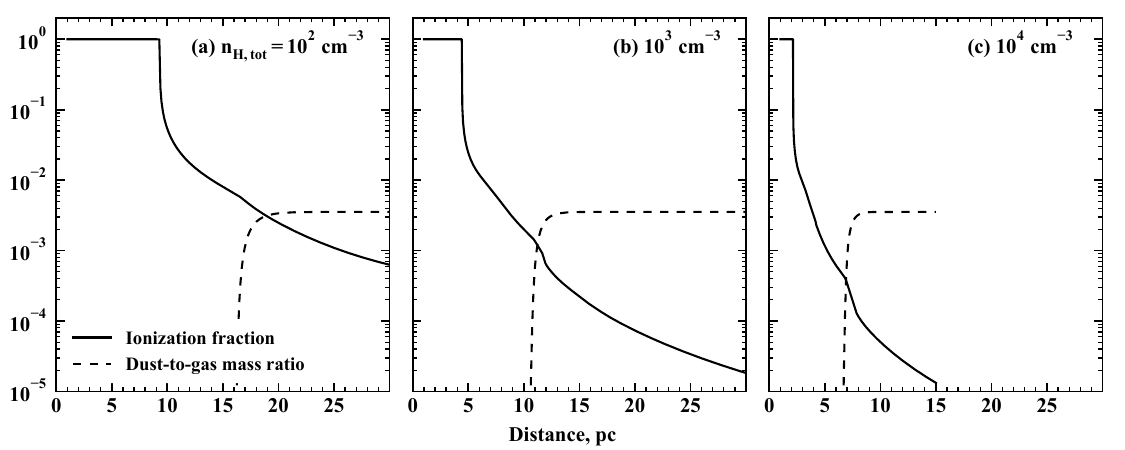}
\caption{\rm The gas ionization fraction and the dust-to-gas mass ratio as a function of the distance. The results of our calculations are presented for three models: (a) model~1, the total number density of hydrogen nuclei in the cloud in this case is $n_{\rm H,tot} = 10^2$~cm$^{-3}$; (b) model~2, $n_{\rm H,tot} = 10^3$~cm$^{-3}$; (c) model~3, $n_{\rm H,tot} = 10^4$~cm$^{-3}$. The numerical simulations of the propagation of the gamma-ray burst radiation were carried out up to distances of 100, 35, and 15~pc in models 1, 2 and 3, respectively.}
\label{fig1}
\end{figure*}

\begin{figure*}
\centering
\includegraphics[width = 0.55\textwidth]{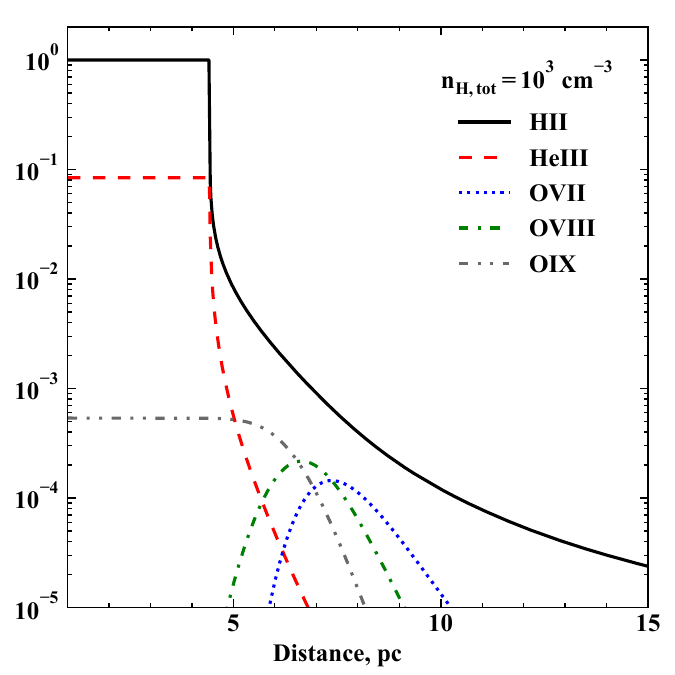}
\caption{\rm The abundances of hydrogen ion, doubly ionized helium, and oxygen ions with a high charge as a function of the distance. The results for the numerical model~2 are shown, the total number density of hydrogen nuclei in the cloud is $n_{\rm H,tot} = 10^3$~cm$^{-3}$.}
\label{fig2}
\end{figure*}

\begin{figure*}
\centering
\includegraphics[width = 1.\textwidth]{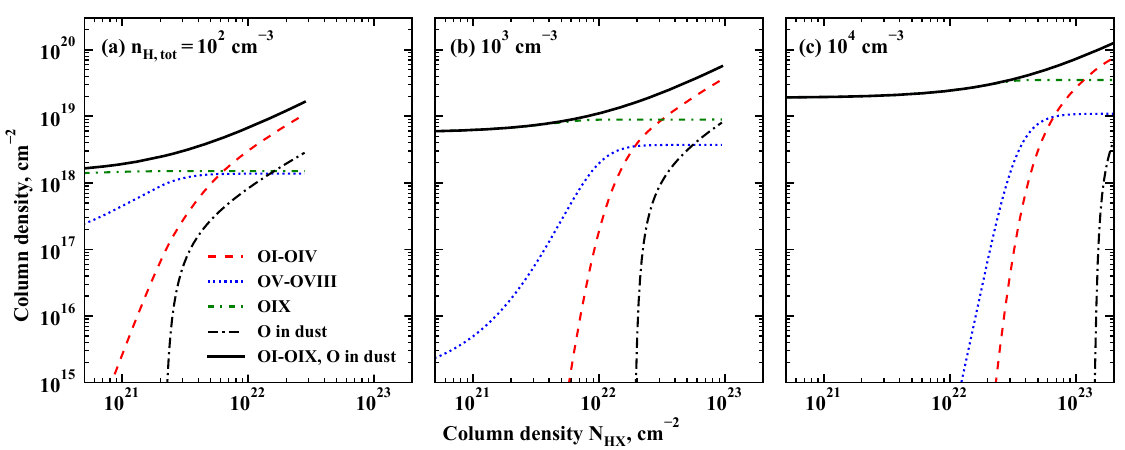}
\caption{\rm The column density of oxygen ions as a function of the hydrogen column density $N_{\rm HX}$ for models 1 (a), 2 (b), and 3 (c). The results are presented for the group of ions with a low charge OI--OIV, the group of ions with a high charge OV--OVIII, the OIX ion without electrons, and oxygen atoms locked in dust grains. The parameter value $N_{\rm HX} = 0$ corresponds to the ionized--neutral gas boundary.}
\label{fig3}
\end{figure*}

\section{Results}
\subsection{The ionization structure of the molecular cloud}
Figure~\ref{fig1} shows the gas ionization fraction and the dust-to-gas mass ratio as a function of the distance from the progenitor star of the gamma-ray burst (at the same retarded time of $10^{5}$~s after the gamma-ray burst start). The results of our calculations are presented for the numerical models 1, 2 and 3 in which the total number density of hydrogen nuclei was assumed to be equal to $n_{\rm H,tot} = 10^2$, $10^3$ and $10^4$~cm$^{-3}$, respectively (see Table~2).
The transition boundary between the region where the gas is completely ionized and the region where the gas is predominantly neutral is sharp -- the gas ionization fraction drops from a value  close to 1 to $\sim 0.1$ at a distance (in units of the hydrogen column density) of $\sim 10^{20}$~cm$^{-2}$. The dust destruction radius $R_{\rm d}$ is about 17, 11, and 7~pc for gas densities $n_{\rm H,tot} = 10^2$, $10^3$ and $10^4$~cm$^{-3}$, respectively.
The hydrogen column density $N_{\rm HX}$ from the ionized--neutral gas boundary to the dust-destruction front is $0.25 \times 10^{22}$, $2.2 \times 10^{22}$, and $16 \times 10^{22}$~cm$^{-2}$ for the three models in question, respectively. 

Figure~\ref{fig2} shows the ion abundances as a function of the distance from the progenitor star of the gamma-ray burst. The results of the calculations are presented for the numerical model~2 ($n_{\rm H,tot} = 10^3$~cm$^{-3}$). A sharp drop in the abundances of hydrogen and helium ions at a distance of about 4.4~pc forms the ionized--neutral gas boundary. However, there is no such sharp boundary for metal ions -- the abundances of metal ions with a high charge decrease gradually with distance. For example, the oxygen is in an ultra-ionized state (OVII--OIX) at a distance up to $7-8$~pc (Fig.~\ref{fig2}).  

The ion column density at a given distance from the gamma-ray burst source is calculated as follows \autocite{Nesterenok2024}: 

\begin{equation}
N_{\rm A}(R_j) = \sum_{i = 1}^{j - 1} n_{\rm A}^i (t_r) \Delta R, 
\end{equation}

\noindent
where $n_{A}^i$ is the number density of ions A in the cloud shell $i$ at the same retarded time $t_r$, and $R_j$ is the inner radius of the cloud shell $j$. The absorption of the gamma-ray burst afterglow at the time $t_r$ after the burst onset is determined by the ion column densities $N_{\rm A}$. 
The most abundant chemical element after hydrogen and helium is oxygen. Figure~\ref{fig3} shows the column density of oxygen ions as a function of the parameter $N_{\rm HX}$ at $t_{\rm r} = 10^5$~s. The photoionization of inner electron shells of metal ions makes the main contribution to the absorption of radiation in the X-ray wavelength range.
The ionization threshold of the electron K-shell of oxygen ions is 538~eV for OI and 871~eV for OVIII \autocite{Verner1995}. The photoionization cross section of the electron K-shell of OVIII ion is approximately half the photoionization cross section of OI ion at photon energies above the ionization threshold of OVIII ion. All OI--OVIII ions contribute to the X-ray absorption.  
For a total number density of hydrogen nuclei in the cloud $n_{\rm H,tot} = 10^2$~cm$^{-3}$ (Fig.~\ref{fig3}a), oxygen is fully ionized in the near-surface layer behind the ionized--neutral gas boundary, the hydrogen column density of this layer is $N_{\rm HX} \approx 10^{20}$~cm$^{-2}$.
For a total number density of hydrogen nuclei in the cloud $n_{\rm H,tot} = 10^4$~cm$^{-3}$ (Fig.~\ref{fig3}c), the hydrogen column density of the neutral cloud layer where oxygen is fully ionized is much larger -- $N_{\rm HX} \approx 2.5 \times 10^{22}$~cm$^{-2}$. In the neutral cloud layer where metal ions are in an ultra-ionized state, helium atoms and hydrogen atoms and molecules make a major contribution to the X-ray absorption.

\begin{table*}
	\caption{\normalsize The results of the fitting of gamma-ray burst afterglow spectra}
	\centering
	\begin{tabular}{p{3cm}|p{2.5cm}|p{2cm}|p{2cm}|p{2cm}|p{2cm}}
	\toprule
Gamma-ray burst & Redshift, $z$ & \multicolumn{4}{c}{$N_{\rm HX}$, $\times 10^{22}$~cm$^{-2}$} \\[3pt]
\cline{3-6}
 & & tbabs & model 1 & model 2 & model 3 \\[3pt]
\midrule
050802  & 1.71 & $0.29_{-0.15}^{+0.16}$ & $0.39_{-0.15}^{+0.15}$ & $0.86_{-0.31}^{+0.23}$ & $1.0_{-0.6}^{+0.6}$ \\[3pt]
060111A & 2.32 & $1.5_{-0.6}^{+0.7}$ & $1.6_{-0.6}^{+0.7}$ & $2.2_{-0.6}^{+0.7}$ & $4.4_{-1.1}^{+1.0}$ \\[3pt]
060306  & 1.55 & $3.8_{-0.6}^{+0.7}$ & $4.0_{-0.7}^{+0.7}$ & $4.5_{-0.6}^{+0.7}$ & $7.3_{-0.7}^{+0.8}$ \\[3pt]
060908  & 1.88 & $1.9_{-1.0}^{+1.3}$ & $1.9_{-1.0}^{+1.4}$ & $2.6_{-1.0}^{+1.3}$ & $4.8_{-1.8}^{+1.7}$ \\[3pt]
071021  & 2.45 & $1.9_{-0.6}^{+0.6}$ & $2.0_{-0.6}^{+0.7}$ & $2.6_{-0.6}^{+0.6}$ & $5.0_{-0.8}^{+0.8}$ \\[3pt]
071117  & 1.33 & $1.7_{-0.5}^{+0.6}$ & $1.7_{-0.5}^{+0.6}$ & $2.4_{-0.5}^{+0.5}$ & $4.4_{-0.8}^{+0.8}$ \\[3pt]
080210  & 2.64 & $2.2_{-1.1}^{+1.3}$ & $2.4_{-1.1}^{+1.3}$  & $2.9_{-1.1}^{+1.3}$ & $5.5_{-1.5}^{+1.5}$ \\[3pt]
080310  & 2.43 & $0.5_{-0.4}^{+0.4}$ & $0.6_{-0.4}^{+0.4}$ & $1.2_{-0.6}^{+0.5}$ & $2.2_{-1.9}^{+1.3}$ \\[3pt]
080805  & 1.50 & $1.8_{-0.8}^{+1.1}$ & $1.8_{-0.8}^{+1.1}$ & $2.5_{-0.7}^{+1.0}$ & $4.8_{-1.5}^{+1.5}$ \\[3pt]
080928  & 1.69 & $0.39_{-0.21}^{+0.22}$ & $0.48_{-0.20}^{+0.21}$ & $1.0_{-0.4}^{+0.3}$ & $1.3_{-0.8}^{+0.8}$ \\[3pt]
081109A & 0.98 & $1.4_{-0.3}^{+0.3}$ & $1.5_{-0.3}^{+0.3}$ & $2.2_{-0.3}^{+0.3}$ & $3.9_{-0.5}^{+0.5}$ \\[3pt]
090424  & 0.54 & $0.58_{-0.10}^{+0.10}$ & $0.70_{-0.09}^{+0.10}$ & $1.16_{-0.13}^{+0.13}$ & $1.5_{-0.3}^{+0.3}$ \\[3pt]
090926B & 1.24 & $2.4_{-1.3}^{+1.9}$ & $2.5_{-1.3}^{+2.0}$ & $3.1_{-1.3}^{+1.9}$ & $5.5_{-2.1}^{+2.3}$ \\[3pt]
091109A & 3.08 & $1.68_{-1.3}^{+1.6}$ & $1.8_{-1.4}^{+1.6}$ & $2.4_{-1.4}^{+1.6}$ & $4.8_{-3.0}^{+1.9}$ \\[3pt]
100621A & 0.54 & $2.75_{-0.3}^{+0.3}$ & $2.8_{-0.3}^{+0.3}$ & $3.4_{-0.3}^{+0.3}$ & $5.8_{-0.4}^{+0.4}$ \\[3pt]
100901A & 1.41 & $0.34_{-0.14}^{+0.15}$ & $0.45_{-0.14}^{+0.14}$ & $0.93_{-0.26}^{+0.21}$ & $1.2_{-0.5}^{+0.5}$ \\[3pt]
110715A & 0.82 & $1.5_{-0.4}^{+0.5}$ & $1.6_{-0.4}^{+0.5}$ & $2.2_{-0.4}^{+0.5}$ & $4.4_{-0.8}^{+0.7}$ \\[3pt]
120118B & 2.94 & $8_{-3}^{+3}$ & $9_{-3}^{+3}$ & $9_{-3}^{+3}$ & $12_{-3}^{+3}$ \\[3pt]
120326A & 1.80 & $0.63_{-0.19}^{+0.20}$ & $0.73_{-0.18}^{+0.20}$ & $1.31_{-0.24}^{+0.23}$ & $2.5_{-0.7}^{+0.5}$ \\[3pt]
121024A & 2.30 & $1.3_{-0.6}^{+0.7}$ & $1.4_{-0.6}^{+0.7}$ & $2.0_{-0.6}^{+0.7}$ & $4.2_{-1.1}^{+1.0}$ \\[3pt]
130420A & 1.30 & $0.49_{-0.19}^{+0.21}$ & $0.6_{-0.18}^{+0.19}$ & $1.1_{-0.3}^{+0.3}$ & $1.6_{-0.6}^{+0.7}$ \\[3pt]
140512A & 0.72 & $0.31_{-0.07}^{+0.07}$ & $0.43_{-0.06}^{+0.07}$ & $0.80_{-0.14}^{+0.12}$ & $0.85_{-0.20}^{+0.21}$ \\[3pt]
140629A & 2.28 & $0.8_{-0.3}^{+0.4}$ & $0.9_{-0.3}^{+0.4}$ & $1.5_{-0.4}^{+0.4}$ & $3.1_{-1.0}^{+0.7}$ \\[3pt]
140907A & 1.21 & $0.8_{-0.4}^{+0.4}$ & $0.9_{-0.4}^{+0.4}$ & $1.5_{-0.5}^{+0.5}$ & $2.9_{-1.3}^{+1.0}$ \\[3pt]
150910A & 1.36 & $0.13_{-0.12}^{+0.12}$ & $0.24_{-0.20}^{+0.13}$ & $0.5_{-0.4}^{+0.3}$ & $0.5_{-0.4}^{+0.4}$ \\[3pt]
151112A & 4.1 & $3.2_{-1.3}^{+1.4}$ & $3.3_{-1.3}^{+1.4}$ & $3.8_{-1.3}^{+1.4}$ & $6.7_{-1.5}^{+1.5}$ \\[3pt]
170604A & 1.33 & $0.15_{-0.12}^{+0.12}$ & $0.27_{-0.18}^{+0.12}$ & $0.5_{-0.4}^{+0.3}$ & $0.5_{-0.4}^{+0.4}$ \\[3pt]
	\bottomrule
	\end{tabular}
	
\vspace{3mm}
\raggedright
The gamma-ray bursts 071117, 090424, 100621A, 110715A, and 140512A enter into the catalogue of Tsvetkova et al. \autocite{Tsvetkova2017}, the remaining gamma-ray bursts enter into the catalogue of Tsvetkova et al. \autocite{Tsvetkova2021}.	
\end{table*}

\begin{figure*}
\centering
\includegraphics[width = 0.53\textwidth]{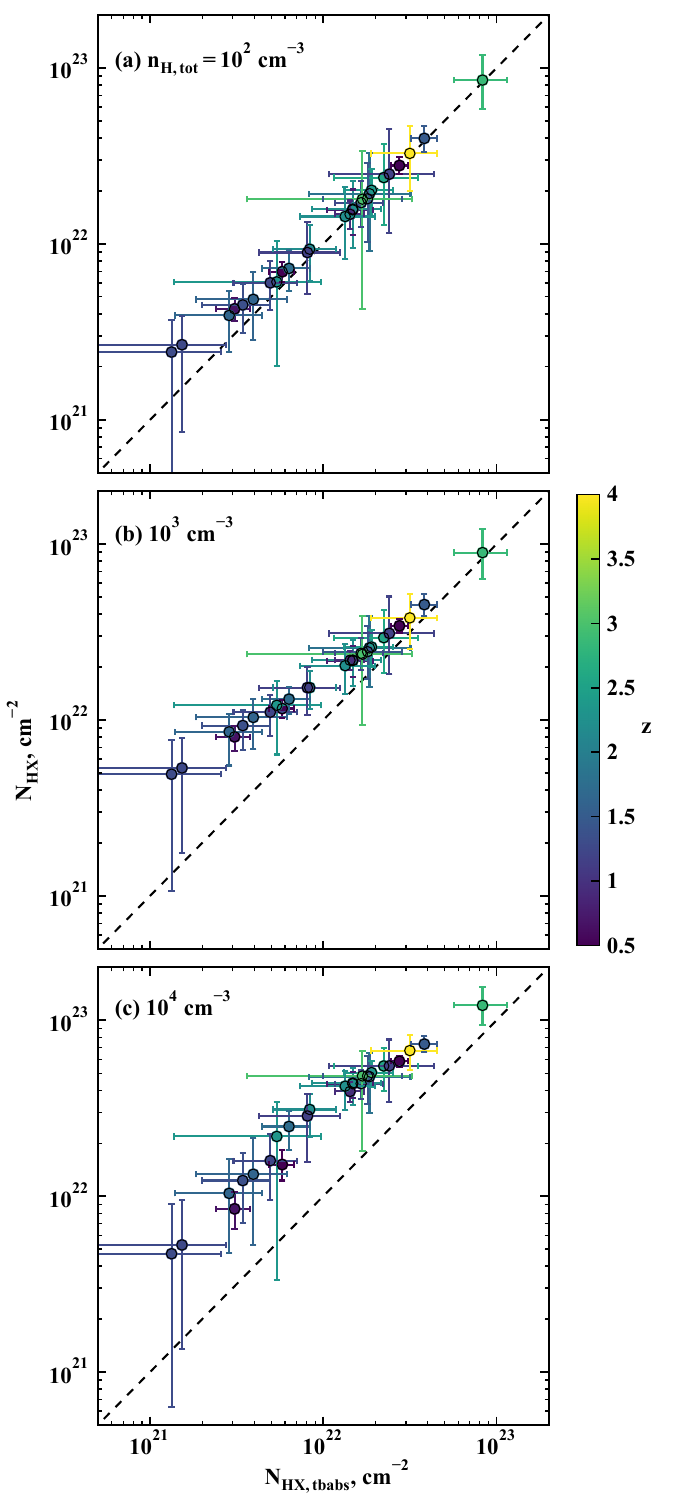}
\caption{\rm The hydrogen column densities evaluated by fitting the energy spectra of 27 gamma-ray bursts from Table~4. The horizontal axis presents values of $N_{\rm HX, tbabs}$ obtained by fitting the spectra using the tbabs model -- the first composite model in (\ref{eq_fit_models}). The vertical axis shows the values of $N_{\rm HX}$ which were calculated based on the second composite model in (\ref{eq_fit_models}), i.e., using the results of our numerical simulations. Panels (a), (b), (c) show the results of the numerical models 1, 2, and 3 from Table~2. The color scale indicates the gamma-ray burst redshift.}
\label{fig4}
\end{figure*}

\subsection{The results of our calculations of $N_{\rm HX}$}
The energy spectra of the afterglows of 45 gamma-ray bursts were fitted in the XSpec program using two composite models (\ref{eq_fit_models}). For 18 gamma-ray bursts, the fitting of the energy spectra provides either the upper limits on $N_{\rm HX}$, or the error in the evaluation of $N_{\rm HX}$ is equal to the most probable value of this parameter. Below we do not provide the results for these gamma-ray bursts.   
For the remaining 27 gamma-ray bursts, the derived values of the hydrogen column density $N_{\rm HX}$ are given in Table~4 and shown in Figure~\ref{fig4}. In Figure~\ref{fig4}, the hydrogen column densities $N_{\rm HX, tbabs}$ calculated using the tbabs model, the first composite model in (\ref{eq_fit_models}), are shown along the horizontal axis. The vertical axis shows the values of $N_{\rm HX}$ calculated using the second composite model in (\ref{eq_fit_models}).
The results of our numerical simulations are used in this model (the numerical models 1, 2, and 3 from Table~2). For a total number density of hydrogen nuclei $n_{\rm H,tot} = 10^2$~cm$^{-3}$, the difference between the results obtained using the two models is significant only for gamma-ray bursts with small values of $N_{\rm HX} \leq 3 \times 10^{21}$~cm$^{-2}$ (Fig.~\ref{fig4}a). In this case, the difference in the values of hydrogen column densities obtained using the two models is smaller than the errors in this parameter.

For a total number density of hydrogen nuclei $n_{\rm H,tot} = 10^4$~cm$^{-3}$, the hydrogen column densities $N_{\rm HX}$ obtained from the results of our numerical simulations significantly exceed $N_{\rm HX, tbabs}$ (Fig.~\ref{fig4}c). The average ratio of column densities $N_{\rm HX, tbabs}/N_{\rm HX}$ is about 0.3 for $N_{\rm HX, tbabs} \leq 2 \times 10^{22}$~cm$^{-2}$.
This value of $N_{\rm HX, tbabs}/N_{\rm HX}$ is approximately equal to the contribution of helium atoms and hydrogen molecules to the X-ray absorption in an unperturbed interstellar molecular gas -- about 0.25 at photon energies $E = 1$~keV. According to our simulations, the metal ions are in an ultra-ionized state behind the ionized--neutral gas boundary and do not contribute noticeably to the absorption of the radiation (Fig.~\ref{fig3}c).
The tbabs model does not take into account the metal ionization, and describes the X-ray absorption in an unperturbed interstellar gas \autocite{Wilms2000}. At a given hydrogen column density in the cloud, the gas is located closer to the radiation source at higher gas densities in the cloud. Therefore, the effect of metal ionization on the X-ray absorption is more pronounced for higher gas densities in the cloud (Fig.~\ref{fig4}c). 

\subsection{The results of our calculations of $N_{\rm HX}$ for gas metallicities $[M/H] = -0.5$ and $-1$}
\begin{figure*}
\centering
\includegraphics[width = 1.\textwidth]{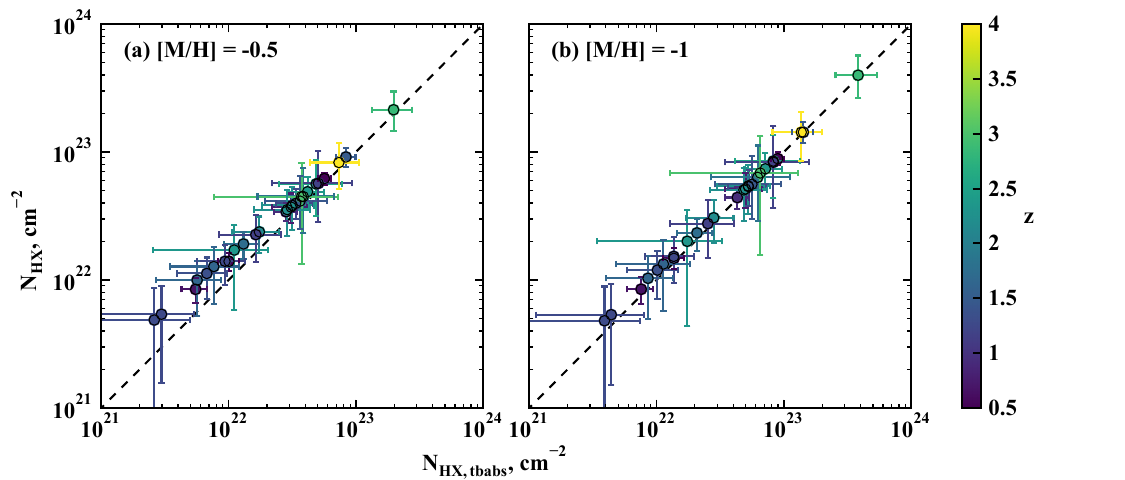}
\caption{\rm The hydrogen column densities calculated by fitting the spectra of gamma-ray bursts. The hydrogen column densities $N_{\rm HX, tbabs}$ obtained using the tbabs absorption model are plotted along the horizontal axis. The hydrogen column densities $N_{\rm HX}$ calculated based on the results of our numerical simulations are plotted along the vertical axis (the numerical models 4 and 5 from Table~3). Panel (a) shows the results for a gas metallicity in the host galaxy $[M/H] = -0.5$, and panel (b) -- for a gas metallicity $[M/H] = -1$.}
\label{fig5}
\end{figure*}

Figure~\ref{fig5} shows the results of the spectrum fitting of the afterglows of 27 gamma-ray bursts listed in Table~4. In the fitting, we used the models based on the results of our numerical simulations (the numerical models 4 and 5 from Table~3) and the tbabs absorption model. The gas metallicity in the host galaxy was assumed to be $[M/H] = -0.5$ (Fig.~\ref{fig5}a) and $-1$ (Fig.~\ref{fig5}b).
For a gas metallicity $[M/H] = -0.5$, the difference between the values of $N_{\rm HX}$ and $N_{\rm HX, tbabs}$ obtained using the two models (\ref{eq_fit_models}), is comparable to the measurement errors of the parameter (Fig.~\ref{fig5}a). For a gas metallicity $[M/H] = -1$, the metal ions make a minor contribution to the absorption of radiation in the X-ray wavelength range \autocite{Nesterenok2024}. As a result, the cloud ionization structure has no effect on the evaluation of the hydrogen column density $N_{\rm HX}$. In this case, both absorption models provide close hydrogen column densities (Fig.~\ref{fig5}b).

\subsection{The results of our calculations of $N_{\rm HX}$ for distances from the gamma-ray burst source to the cloud $R_{\rm min}$ = 10 and 100~pc}
\begin{figure*}
\centering
\includegraphics[width = 1.\textwidth]{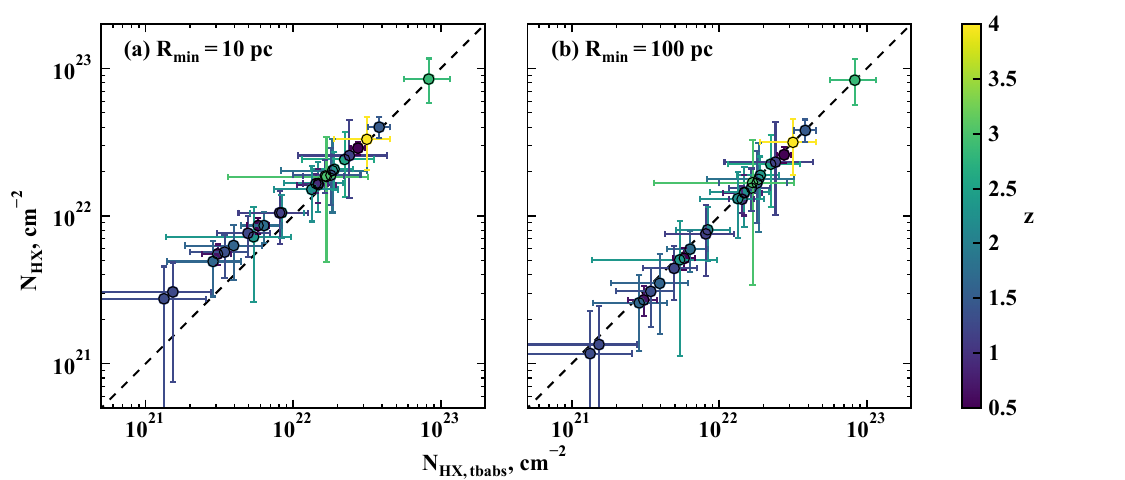}
\caption{\rm The hydrogen column densities calculated by fitting the spectra of gamma-ray bursts. The hydrogen column densities $N_{\rm HX}$ calculated based on the results of our numerical simulations are plotted along the vertical axis: (a) the results of our calculations in which the distance between the gamma-ray burst source and the cloud is $R_{\rm min}$ = 10~pc; (b) the results of our calculations in which $R_{\rm min}$ = 100~pc (the numerical models 6 and 7 from Table~3). The gas metallicity is solar, $[M/H] = 0$.}
\label{fig6}
\end{figure*}

In this section we present the results of the spectrum fitting of gamma-ray burst afterglows using the results of our numerical simulations in which the distance from the gamma-ray burst source to the cloud was taken equal to $R_{\rm min} = 10$ and $100$~pc (the numerical models 6 and 7 from Table~3). The results of the calculations of the hydrogen column densities are presented in Fig.~\ref{fig6}.
The gas metallicity in the X-ray absorption models, i.e., in the models based on the results of numerical simulations and the tbabs model, was assumed to be equal to the solar metallicity. The gas layer behind the ionization front in which the metal ions are in an ultra-ionized state shortens with increasing distance from the gamma-ray burst source. Already at $R_{\rm min} = 10$~pc the difference between the values of $N_{\rm HX}$ and $N_{\rm HX, tbabs}$ is comparable to the measurement errors of the parameter.

\section{Discussion}
\subsection{The gas metallicity in gamma-ray burst host galaxies}
The gamma-ray bursts at low redshifts $z < 1$ are typically produced in faint low-mass star-forming galaxies (see, e.g., \autocite{Vergani2015}). The fraction of gamma-ray bursts in massive galaxies with a high overall star formation rate increases with increasing redshift \autocite{Greiner2015,Perley2016}. This observational fact is considered as a consequence of the suppression of the specific formation rate of gamma-ray bursts for a metallicity of the interstellar medium above a certain threshold value \autocite{Kruhler2018}.
The gas metallicity in a galaxy can be estimated from the observations of emission lines of metal ions. The estimates of this  threshold metallicity based on the observations of host galaxies of gamma-ray bursts provide $(0.5-1) \times Z_{\odot}$ \autocite{Perley2016,Graham2017,Vergani2017,Palmerio2019}. The gas metallicity in the host galaxy of a gamma-ray burst can also be derived from the analysis of absorption lines of metal ions in the afterglow.
This method provides an average value of the gas metallicity $Z \sim 0.1 Z_{\odot}$ in the host galaxies of gamma-ray bursts at redshifts $z = 3.5-6$ \autocite{Cucchiara2015}. As a rule, this method yields lower estimates of the gas metallicity than follows from the emission line analysis of gamma-ray burst galaxies, e.g., \autocite{Graham2023}. It should be noted that the gas metallicity in the part of the galaxy where the gamma-ray burst occurred may differ from the average gas metallicity in the galaxy (see, e.g., \autocite{Kruhler2017}).

For gamma-ray bursts at low redshifts ($z \lesssim 1$), the gas metallicity in the host galaxy can be close to the solar metallicity. If a dense molecular cloud is located close to the source of the gamma-ray burst, then the metal ions are ionized to a high charge in the neutral layer of the cloud behind the ionization front.
The hydrogen column density of this near-surface layer depends on the gas density in the cloud and is equal to $N_{\rm HX} \approx 3 \times 10^{22}$~cm$^{-2}$ for a gas density $n_{\rm H,tot} = 10^4$~cm$^{-3}$. In this cloud layer, the metal ions do not contribute noticeably to the X-ray absorption even for solar metallicity.
The photoionization of helium atoms and hydrogen atoms (molecules) is responsible for the absorption of X-rays. In this case, the absorption model that does not take into account the ionization of metal ions, predicts a value of $N_{\rm HX}$ three times smaller than the actual value (Fig.~\ref{fig4}). At the same time, the absorption model based on our numerical simulations and the tbabs model provide identical estimates of the hydrogen column density in the absorbing cloud for a gas metallicity $[M/H]= -1$ (Fig.~\ref{fig5}).

\subsection{The observation of a change in $N_{\rm HX}$ for some gamma-ray bursts}
The radiation of the gamma-ray burst ionizes the gas located near the progenitor star of the burst. As a result, the afterglow radiation emitted at late times will encounter less absorbing gas on its path than will the radiation emitted at earlier times. In the case of a dense compact cloud, this change in the hydrogen column density of the absorbing cloud can be observed by analyzing the change in the afterglow energy spectra with time \autocite{Lazzati2002,Valan2023}. 
Valan et al. \autocite{Valan2023} studied the change in the hydrogen column density $N_{\rm HX}$ for 199 gamma-ray bursts with measured spectroscopic redshifts. Valan et al. \autocite{Valan2023} analyzed the Swift/XRT data obtained at the early afterglow stage in the WT mode (at times $t < 200$~s in the rest frame).
According to the results by Valan et al. \autocite{Valan2023}, only 7 of the 199 gamma-ray bursts have evidence for a decrease in the hydrogen column density $N_{\rm HX}$. One gamma-ray burst from this list falls into our sample -- GRB~090926B at redshift $z = 1.24$. The tbabs model predicts that the hydrogen column density in the absorbing cloud is equal to $2.4 \times 10^{22}$~cm$^{-2}$ (for solar metallicity). The absorption model based on the results of our numerical simulations predicts a parameter value of $5.5 \times 10^{22}$~cm$^{-2}$ for a gas density in the cloud $n_{\rm H,tot} = 10^4$~cm$^{-3}$, see Table~4. 
Note that gamma-ray burst GRB~090926B belongs to dark gamma-ray bursts, the visual extinction for this gamma-ray burst is $A_{\rm V} = 1.42^{+1.08}_{-0.57}$ \autocite{Greiner2011}. Most of the gamma-ray bursts have no evidence for a decrease in $N_{\rm HX}$ at the early afterglow stage according to the results by Valan et al. \autocite{Valan2023}. This may be a consequence of the fact that the interstellar gas that absorbs the radiation is not located compactly near the progenitor star of the gamma-ray burst, but lies at a distance or is distributed over a large region in the host galaxy. 

\subsection{The absorption of afterglow radiation in the optical wavelength range}
The detection of afterglow radiation in the infrared and optical wavelength ranges, along with X-ray radiation, allows one to jointly fit the optical and X-ray energy spectra \autocite{Greiner2011,Covino2013}. The hydrogen column densities that follow from the estimates of visual extinction $A_{\rm V}$, are, as a rule, an order of magnitude smaller than the values of $N_{\rm HX}$, i.e., the absorption of X-rays is much stronger.
In such estimates, the solar metallicity is usually adopted, while the dust parameters correspond to the local group of galaxies. If the gamma-ray burst radiation passes through a dense molecular cloud, then the dust-destruction radius is a factor of $2-3$ larger than the ionization front radius (Fig.~\ref{fig1}).
In our model, the characteristic time of the dust grain sublimation is about $3-5$~s from the gamma-ray burst onset (the absorption of UV radiation of the optical flash of the gamma-ray burst leads to the heating and thermal sublimation of dust grains). The metal atoms are liberated from the dust into the gas phase and are ionized to a high charge by the succeeding gamma-ray burst radiation. Thus, one of the reasons for the relatively weak absorption of the optical continuum radiation may be the destruction of dust grains.

\section{Conclusions}
We presented the results of our numerical simulations of the propagation of the gamma-ray burst radiation in a dense molecular cloud for various gas densities, metallicities, and distances from the progenitor star of the gamma-ray burst to the cloud. If the cloud is close to the progenitor star of the gamma-ray burst, $R_{\rm min} = 1$~pc, the metal ions are ionized to a high charge in the cloud layer behind the ionization front. 
For a gas density $n_{\rm H,tot} = 10^4$~cm$^{-3}$, the hydrogen column density of this cloud layer is $N_{\rm HX} \approx 3 \times 10^{22}$~cm$^{-2}$. For such hydrogen column densities in the cloud, the ionization of helium atoms, hydrogen atoms and molecules makes a major contribution to the absorption of the afterglow in the X-ray wavelength range -- irrespective of the gas metallicity.

We considered a sample of 45 long gamma-ray bursts with known redshifts in which the total isotropic equivalent gamma-ray energy $E_{\gamma, {\rm iso}}$ lies in the range from $3.33 \times 10^{52}$ to $7.5\times 10^{52}$~erg. For these gamma-ray bursts we analyzed the energy spectra of their afterglows obtained with the Swift/XRT at late times, $t \geq 4 \times 10^3$~s.
We fitted the energy spectra using the absorption model based on the numerical simulations, that takes into account the gas ionization by the gamma-ray burst radiation, and the tbabs absorption model \autocite{Wilms2000}. We showed that the energy spectrum fitting using the tbabs model could lead to the hydrogen column densities that are smaller than the actual values approximately by a factor of 3 -- if the gas--dust cloud, that absorbs the radiation, is located near the source of the gamma-ray burst.
In particular, the tbabs absorption model provides a hydrogen column density of $2.4 \times 10^{22}$~cm$^{-2}$ for GRB~090926B, whereas the absorption model based on the results of our numerical simulations predicts $5.5 \times 10^{22}$~cm$^{-2}$ for a gas density in the cloud $n_{\rm H,tot} = 10^4$~cm$^{-3}$.

The absorption model based on the results of our numerical simulations and the tbabs model predict the identical values of the hydrogen column density if the molecular cloud is located at a distance $R_{\rm min} > 10$~pc from the gamma-ray burst source or the gas metallicity is $[M/H] = -1$. Our results should be taken into account in the analysis of the energy spectra of gamma-ray bursts whose radiation passed through a dense gas--dust cloud near the progenitor star of the gamma-ray burst -- for example, if a decrease in the hydrogen column density $N_{\rm HX}$ is observed.

\section*{Acknowledgements}
I am grateful to D. Svinkin for his help in processing the observational data.


\printbibliography 

\end{document}